\documentclass[9pt,conference]{IEEEtran}

\usepackage[preprint]{waspaa25}

\usepackage{bm} 

\usepackage{graphicx}

\usepackage{color}
\usepackage{booktabs}
\usepackage[style=base]{caption}
\usepackage{subfigure}
\usepackage{subcaption}
\usepackage{tikz}
\usetikzlibrary{shapes.geometric, arrows.meta, positioning}

\tikzset{
    process/.style={rectangle, minimum width=3cm, minimum height=1cm, text centered, draw=black},
    arrow/.style={->, thick, >=Stealth}
}
\usepackage{pgfplots}
\pgfplotsset{compat=1.18}

\title{Temporal Adaptation of Pre-trained Foundation Models for Music Structure Analysis}

\name{Yixiao Zhang,
      Haonan Chen,
      Ju-Chiang Wang,
      Jitong Chen}
\address{ByteDance}

\begin{document}

\maketitle

\begin{abstract}
Audio-based music structure analysis (MSA) is an essential task in Music Information Retrieval that remains challenging due to the complexity and variability of musical form. Recent advances highlight the potential of fine-tuning pre-trained music foundation models for MSA tasks. However, these models are typically trained with high temporal feature resolution and short audio windows, which limits their efficiency and introduces bias when applied to long-form audio. This paper presents a temporal adaptation approach for fine-tuning music foundation models tailored to MSA. Our method enables efficient analysis of full-length songs in a single forward pass by incorporating two key strategies: (1) \emph{audio window extension} and (2) \emph{low-resolution adaptation}. Experiments on the Harmonix Set and RWC-Pop datasets show that our method significantly improves both boundary detection and structural function prediction, while maintaining comparable memory usage and inference speed. \footnote{Project page: \url{https://sites.google.com/view/temporal-adaptation-for-msa/}}

\end{abstract}

\section{Introduction}

Music exhibits structural complexity across multiple temporal levels~\cite{gttm, jiang2022}, from fine-grained metrical patterns~\cite{ellis2007beat,mcfee2014beat,bock2016joint} to the overarching organization of musical ideas~\cite{buisson2024using}. Analyzing these hierarchical structures remains a central challenge in Music Information Retrieval (MIR), with methodologies evolving from early approaches such as self-similarity matrices and clustering~\cite{foote2000automatic, levy2008structural, nieto2013convex, nieto2014music} to more recent approaches based on deep learning~\cite{ullrich2014boundary, wang2021supervised, wang2021deepstruc, tocatch, buisson2024using, buisson2022learning}.
This work focuses on the task of music structure analysis (MSA)~\cite{mirexstructure, smith2013meta, survey, intro1, intro2, intro3, paulus2010state}, which involves segmenting a music recording into distinct sections and assigning structural function labels (e.g., verse, chorus, and bridge) to these segments.

While recent progress in pre-trained audio foundation models have advanced various MIR tasks~\cite{mert, MusicFM, MuQ, ma2024foundation, yuan_marble_2023, jukemir} and fine-tuning these models for structural segmentation has shown promising results~\cite{buisson2024using, MuQ, MusicFM}, their use in MSA exposes architectural limitations.
Most existing approaches fine-tune pre-trained models by adding task-specific layers, while preserving original input configurations (e.g., sampling rate, audio window size, and temporal resolution).
However, this paradigm presents two major mismatches for MSA: (1) the standard 30-second input windows used during pre-training are too short to capture the long-range dependencies needed for structural analysis; and (2) the fixed high temporal resolution (e.g., 25 Hz) leads to excessive local details, increasing computational cost without improving global structure modeling~\cite{buisson2022learning}.
Together, these mismatches hinder the alignment between model capabilities and the functional requirements of MSA. 


\begin{figure}[tbp]
    \centering
    \includegraphics[trim=25pt 25pt 20pt 35pt,clip,width=0.8\linewidth]{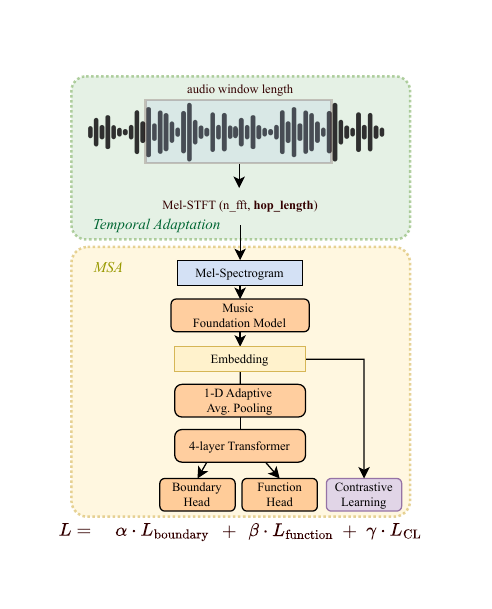}
    \caption{An overview of our proposed temporal adaptation method. Built on a pre-trained Music Foundation Model, our approach adjusts both the audio window length and the temporal resolution of Mel-Spectrogram during fine-tuning, enabling the model to process significantly longer audio inputs.}
    \label{fig:diagram}
\end{figure}

\begin{figure*}[tbp]
    \centering
    \includegraphics[trim=33pt 0pt 35pt 0,clip,width=\linewidth]{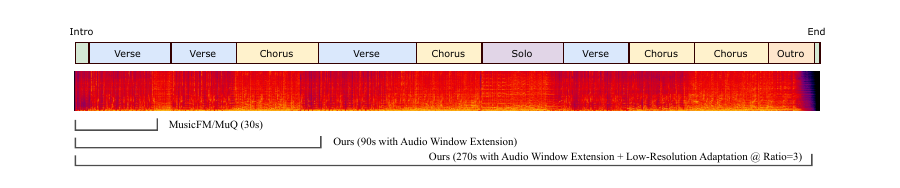}
    \caption{Comparison of audio window lengths across different methods. Fine-tuning with temporal adaptation enables processing full-song windows (e.g., 270s), enhancing the model to capture long-range dependencies. The spectrogram example is selected from the Harmonix Set.}
    \label{fig:teaser}
\end{figure*}

In this paper, we propose \textbf{temporal adaptation}, a fine-tuning method that adjusts both the temporal resolution and audio input length to address key limitations in applying pre-trained models to MSA. Our approach consists of two complementary strategies: (1) \emph{Audio Window Extension}, which enables the model to process audio inputs longer than those used during pre-training; and (2) \emph{Low-Resolution Adaptation}, which strategically reduces the temporal resolution of feature extraction by a fixed ratio (e.g., 3$\times$). Together, these strategies substantially expand the model’s receptive field, enhancing its ability to capture long-range structural dependencies while improving computational efficiency. Controlled experiments show that, under equivalent memory constraints, our method allows the model to process audio up to 9 times longer (see Figure~\ref{fig:teaser}) and improves prediction accuracy (see Table~\ref{tab:main_results_cross_val} and \ref{tab:main_results_cross_dataset}).


Our evaluation on the Harmonix Set~\cite{harmonix} and RWC-Pop~\cite{rwc} datasets reveals three key findings. First, full-length song processing yields an absolute improvement of approximately 10\% in section labeling accuracy compared to window-based baselines. Second, systematic ablation studies across audio window lengths (24 to 90 seconds) and temporal resolution reduction ratios (1$\times$ to 5$\times$) identify optimal trade-offs between accuracy and efficiency. Third, combining extended audio input with reduced temporal resolution produces more temporally consistent boundary detection. In particular, this strategy significantly increases prediction accuracy without additional memory usage or more inference time than baselines.

\section{Methodology}
\label{sec:method}

\subsection{System Overview}
Figure~\ref{fig:diagram} illustrates our MSA system. The input is an audio window of length $T$ seconds, from which frame-based features are extracted using a Mel-scaled Short-Time Fourier Transform (Mel-STFT) with an FFT size of n\_fft and a hop size of hop\_length $h$ (in milliseconds). This produces a Mel-Spectrogram with a temporal resolution of ($1000/h$) frames per second.
Our Music Foundation Model adopts a 12-layer Conformer encoder~\cite{gulati2020conformer} as the backbone, following architectures such as MusicFM~\cite{MusicFM}. This backbone is pre-trained and integrated into our system for fine-tuning on the MSA task. To further enhance the model capacity, we append four additional Transformer layers on top of the foundation model. Consistent with prior work~\cite{tocatch}, the system includes two separate output heads for predicting structural \emph{boundaries} and \emph{functions}, respectively.

\subsection{Temporal Adaptation}

\subsubsection{Audio Window Extension}
\label{sec:length_extension}
During fine-tuning, we extend the audio window length $T$ (e.g., 90s) significantly beyond the duration used in pre-training (e.g., 30s). The motivation is to leverage the inherent generalization capabilities of Transformer architectures to model longer-range dependencies than those seen during pre-training.
While this extension improves the model’s ability to capture global structure, it introduces two key limitations. First, large discrepancies between pre-training and fine-tuning input lengths can introduce a transfer learning gap, hindering effective knowledge transfer and potentially degrading performance, as evidenced by our ablation results (see Figure~\ref{fig:length_ablation}). Second, the self-attention mechanism in Transformers has quadratic time and memory complexity ($O(T^2)$), which imposes practical constraints on input length. In our setup, this limits the maximum feasible window length to about 90 seconds, falling short of the full duration of typical songs (often exceeding 210 seconds).



\subsubsection{Low-Resolution Adaptation}
\label{sec:low-resolution_adaptation}
Pre-trained models are commonly trained with relatively high temporal feature resolutions using shorter hop sizes (e.g., 10ms -- 40ms). While such resolutions are effective for capturing fine-grained audio characteristics, they are often suboptimal for MSA; Moreover, high temporal resolution restricts the maximum input duration that can be processed under memory constraints.
To address this, we propose reducing the temporal resolution during fine-tuning by increasing the hop size used in Mel-spectrogram extraction without altering the model architecture. Specifically, if the pre-training hop\_length is $h$, we set the fine-tuning hop\_length to $N \cdot h$, where $N > 1$ is an integer down-sampling ratio (e.g., $N=3$ corresponds to tripling hop\_length compared to pre-training).


Our Low-Resolution Adaptation method reduces the temporal length of Mel-Spectrogram by a factor of $N$, enabling the model to process an audio window that is $N$ times longer while maintaining a similar computational cost to pre-training. For instance, processing a 90-second audio window at one-third the temporal resolution (i.e., $N=3$) incurs a comparable cost to processing a 30-second window at full resolution.
During fine-tuning, the model initializes from pre-trained weights and learns to adapt its feature representations to this coarser temporal granularity. Although some fine-grained temporal details are inevitably lost, the significantly expanded receptive field allows the model to better capture long-range dependencies, which is an essential aspect for accurate MSA \cite{survey}.

\subsection{Contrastive Learning}\label{sec:CL}

We incorporate an auxiliary contrastive loss during training to serve as a regularizer. As our temporal adaptation method enables the model to process full-length audio of a song, the contrastive loss is able to encourage overall structural consistency by promoting similarity between embeddings from the same label of sections (e.g., verse–verse, chorus–chorus) while pushing apart embeddings from different sections (e.g., verse–chorus, chorus–bridge). 

Formally, let $\mathbf{z}_i$ and $\mathbf{z}_j$ denote the average embeddings of two segments and $I$ indicates whether they share the same label ($I_{ij}=1$) or not ($I_{ij}=0$):
\begin{equation} \label{eq:CL_loss_short}
\ell(\mathbf{z}_i, \mathbf{z}_j) = I_{ij} \cdot \text{dist}(\mathbf{z}_i, \mathbf{z}_j)^2 + (1 - I_{ij}) \cdot \max(0, m - \text{dist}(\mathbf{z}_i, \mathbf{z}_j))^2
\end{equation}
where $\text{dist}(\cdot,\cdot)$ is the Euclidean distance and $m=1$ is margin. The total contrastive loss $\mathcal{L}_{\text{CL}}$ is the average of $\ell$ over selected pairs within a training song. Embeddings are linearly projected to a lower-dimensional space prior to distance computation.

\subsection{Learning Objective}

All time-based annotations are converted into frame-level activation curves for both boundary and structural function labels, following the procedure in \cite{tocatch}.

For the \emph{structural function} head, we use a weighted binary cross-entropy (WBCE) loss:
\begin{equation}
    \mathcal{L}_{\text{function}} = - \sum_{i} w_i \left[ y_i \log \hat{y}_i + (1 - y_i) \log (1 - \hat{y}_i) \right],
\label{eq_wbce}
\end{equation}
where $w_i$ is a class-balancing weight, and $y_i$ and $\hat{y}_i$ denote the ground-truth and predicted values, respectively.

For the \emph{boundary detection} head, training is more challenging due to the extreme sparsity of boundary activations. To address this, we adopt a composite loss:
$ \mathcal{L}_{\text{boundary}} =  \mathcal{L}_{\text{WBCE}} + \mathcal{L}_{\text{smooth-L1}} + \mathcal{L}_{\text{focal}} $
where $\mathcal{L}_{\text{WBCE}}$ follows the same form as Equation~\ref{eq_wbce}, L1 smooth loss $\mathcal{L}_{\text{smooth-L1}}$ is used for robustness against outliers, and focal loss $\mathcal{L}_{\text{focal}}$~\cite{lin2017focal} mitigates the class imbalance by emphasizing hard examples.


Since the three loss terms may differ in scale, we normalize each by its magnitude and combine them with fixed weights:
\begin{equation}
    \mathcal{L} = \alpha \frac{\mathcal{L}_{\text{boundary}}}{\|\mathcal{L}_{\text{boundary}}\|} + \beta \frac{\mathcal{L}_{\text{function}}}{\|\mathcal{L}_{\text{function}}\|} + \gamma \frac{\mathcal{L}_{\text{CL}}}{\|\mathcal{L}_{\text{CL}}\|},
\end{equation}
where $\alpha$ = 0.1, $\beta$ = 0.9, and $\gamma$ = 0.1 are empirically selected weights. 


\section{Experiments}

\subsection{Datasets}

We use two public datasets for the experiments. \textbf{Harmonix}~\cite{harmonix} contains 912 annotated tracks spanning multiple genres such as pop, rock, electronic, and hip-hop. \textbf{RWC-Pop}~\cite{rwc} consists of 100 annotated tracks from the RWC dataset’s popular music subset. 
Following prior work, we conduct both cross-validation and cross-dataset evaluations. For cross-validation, we adopt the 8-fold setup on Harmonix Set, which is consistent with our main baselines~\cite{allinone,buisson2024using}. For cross-dataset evaluation, we train on one dataset and evaluate on the other. In both settings, a portion of the training data is held out as a validation set for model selection during training.


\subsection{Metrics}

The following metrics are used to measure the performance of segment boundary detection and structural function prediction. (1) \textbf{HR.5F}: F-measure of boundary hit rate at 0.5 seconds; (2) \textbf{HR3F}: F-measure of boundary hit rate at 3 seconds; (3) \textbf{Accuracy (ACC)}: the frame-wise accuracy between the predicted function and the ground-truth label.

\subsection{Baselines}
We compare our model against both pre-trained and non-pre-trained models. Among the pre-trained baselines, we consider MERT~\cite{mert} and MusicFM~\cite{MusicFM}, which are music foundation models that support fine-tuning with structural boundary and function prediction heads. Given that MusicFM reported its MSA results on a fixed pre-split dataset, we took its official checkpoint and re-trained its probing head with cross-validation and cross-dataset data splits.


For non-pre-trained methods, we include Harmonic-CNN~\cite{won2019automatic}, SpecTNT~\cite{specTNT}, and All-In-One~\cite{allinone}, all of which are trained from scratch. We also consider the model by {Morgan et al.}~\cite{buisson2024using}, which leverages a graph neural network over a pre-trained frame encoder to capture long-range structure in music. This method represents the current state-of-the-art. 
SpecTNT is evaluated on the Harmonix Set using 4-fold cross-validation, along with additional public datasets. Both All-In-One and the model of {Morgan et al.} are evaluated exclusively on the Harmonix Set using 8-fold cross-validation. We exclude MuQ~\cite{MuQ} from our comparison due to the use of an undisclosed split of Harmonix Set during fine-tuning, which introduces potential unfairness in comparison to other models.



\subsection{Experimental Settings}

We employed MusicFM as our backbone, using its Mel-spectrogram settings (n\_fft=2048, mel\_band=128, hop\_length=240) and a (1024, 64) linear probing layer for contrastive learning computation.

During fine-tuning, we adopt mixed-precision training (fp-16) with a batch size of 8, and a learning rate of 1e-5 with an exponential scheduler that decays every 500 steps with a gamma of 0.99. All models were trained for 75k steps, at which point we observed convergence across experiments. Training was conducted on a single NVIDIA A100-80GB GPU, with each model taking approximately 8 to 12 hours to complete under this configuration.
    


For data preprocessing, we randomly crop each training song to the specified input length, applying zero-padding when the audio is shorter than the target window. To facilitate learning of boundaries and structural functions, we use the label-to-activation technique from~\cite{tocatch}, which smooths hard labels using a Hamming window ramp. For post-processing, we apply the peak-picking algorithm~\cite{ullrich2014boundary} to detect the boundaries.


\subsection{Main Results}
To identify effective configurations for our main experiments, we conducted a grid search over various combinations of audio window lengths $T$ and down-sampling ratio $N$ on a fixed split of the Harmonix Set. From this search, we selected four representative configurations for cross-validation: 100s with $N$=2, 150s with $N$=3, 180s with $N$=2, and 270s with $N$=3.

Table~\ref{tab:main_results_cross_val} summarizes the cross-validation results on Harmonix Set. Our models achieve state-of-the-art performance in terms of ACC and HR3F. Compared to the original MusicFM without temporal adaptation, our method yields statistically significant improvements in both metrics (paired t-test, $p < 0.05$). In cross-dataset evaluation (Table~\ref{tab:main_results_cross_dataset}), our models continue to demonstrate competitive or state-of-the-art performance.
Notably, our temporal adaptation approach reduces fine-grained temporal resolution, which helps scale to longer inputs but may negatively impact performance on stricter metrics such as HR.5F. We further study this trade-off in Section~\ref{sec:down-sampling}.

\begin{table}[tbp]
\footnotesize
    \centering
    \begin{tabular}{l|ccc}
    \toprule
         & ACC & HR.5F & HR3F \\
    \midrule
    \multicolumn{4}{c}{\textbf{Cross-validation Ablation Study}}\\
    \midrule
    & \multicolumn{3}{c}{\textit{Harmonix Set}}\\
    Harmonic-CNN$^\dag$ & 0.680$^{\star}$ & 0.559$^{\star}$ & - \\
    SpecTNT (24s)$^\dag$  & 0.701$^{\star}$ & 0.570$^{\star}$ & - \\
    SpecTNT (36s)$^\dag$   & 0.723$^{\star}$ & 0.558$^{\star}$ & - \\
    All-in-one     & - & \textbf{0.660}$^{\star}$ & - \\
    MERT (5s) & 0.574$^{\star}$ & 0.626$^{\star}$ & - \\
    MusicFM (30s) & 0.725 & 0.640 & 0.729 \\
    Morgan et al.  & 0.742$^{\star}$ & 0.568$^{\star}$ & 0.717$^{\star}$ \\
    \midrule
    Ours [100s ($N$=2), CL] & 0.781  & 0.599 & \textbf{0.828} \\
    Ours [150s ($N$=3), CL] & \textbf{0.787}  & 0.610 & 0.801 \\
    Ours [180s ($N$=2), CL] & 0.781  & 0.601 & 0.826 \\
    Ours [270s ($N$=3), CL] & 0.767 &  0.579 & 0.768 \\
    
    \bottomrule
    \end{tabular}
    \caption{The cross-validation results on Harmonix Set. Scores with a star ($^\star$) represent the reported results from their original papers. Models with a dag ($^\dag$) were evaluated with 4-fold cross-validation.}
    \label{tab:main_results_cross_val}
\end{table}

\begin{table}[tbp]
\footnotesize
    \centering
    \begin{tabular}{l|ccc}
    \toprule
         & ACC & HR.5F & HR3F \\
    \midrule
    \multicolumn{4}{c}{\textbf{Cross-dataset Evaluation}}\\
    \midrule
    & \multicolumn{3}{c}{\textit{Harmonix Set}}\\
    MusicFM & 0.495 &  \textbf{0.597} &{0.695} \\
    Morgan et al.  & {0.530}$^{\star}$& 0.462$^{\star}$ &  0.664$^{\star}$  \\
    \midrule
    Ours [100s ($N$=2), CL] & 0.650 & 0.598 & \textbf{0.803} \\
    Ours [150s ($N$=3), CL] &  0.647 & 0.587 &  0.785 \\
    Ours [180s ($N$=2), CL]   & \textbf{0.652} & 0.589 &  0.786  \\
    Ours [270s ($N$=3), CL] & 0.645 & 0.556 & 0.762\\
    \midrule
    & \multicolumn{3}{c}{\textit{RWC-Pop}}\\
    Harmonic-CNN & 0.646$^{\star}$ & 0.571$^{\star}$ & - \\
    SpecTNT (24s)  & 0.675$^{\star}$ & 0.623$^{\star}$ &  -  \\
    MusicFM &0.680 & 0.636 & 0.764\\
    
    Morgan et al.  & {0.747}$^{\star}$ & 0.648$^{\star}$ &  {0.786}$^{\star}$  \\
    \midrule
    Ours [100s ($N$=2), CL] & 0.748 & 0.534  & \textbf{0.788} \\
    Ours [150s ($N$=3), CL]     & \textbf{0.779} & 0.506 &  0.691  \\
    Ours [180s ($N$=2), CL]     & {0.776} & 0.529 &  {0.765}  \\
    Ours [270s ($N$=3), CL] & 0.762 & 0.482 & 0.667\\
    \bottomrule
    \end{tabular}
    \caption{Cross-dataset structure analysis results on the Harmonix and RWC-Pop datasets. Scores with a star ($^{\star}$) corresponds to the reported results from their original papers.}
    \label{tab:main_results_cross_dataset}
\end{table}

\subsection{Ablation Study}

\subsubsection{Audio Window Length}\label{sec:length}
\begin{figure}[tbp]
    \centering
    \begin{subfigure}
        \centering
        \begin{tikzpicture}
        \begin{axis}[
            xlabel={Audio window length ($T$=24s--90s) with $N$=1},
            xtick=data,
        symbolic x coords={
            {24s},{30s},{40s},{50s},{60s},{70s},{80s},{90s}},
            x tick label style={font=\small},
            y tick label style={font=\small},
            width=\linewidth,
            height={0.4\linewidth},
            grid=both,
            grid style={dashed},
            major grid style={dashed},
            tick label style={font=\small},
            label style={font=\small},
            title style={font=\small},
            legend style={at={(1,1.3)}, legend columns=1}
        ]
    \addplot[orange, mark=*] coordinates {
        ({24s}, 0.742)
        ({30s}, 0.752)
        ({40s}, 0.757)
        ({50s}, 0.774)
        ({60s}, 0.779)
        ({70s}, 0.773)
        ({80s}, 0.773)
        ({90s}, 0.776)
    };
        \addlegendentry{ACC};
        \end{axis}
        \end{tikzpicture}
        \label{fig:down-sampling_acc_1}
    \end{subfigure}
    
    \vskip\baselineskip
    \vspace{-0.5cm}
    
    \begin{subfigure}
        \centering
        \begin{tikzpicture}
        \begin{axis}[
            xlabel={Audio window length ($T$=24s--90s) with $N$=1},
            xtick=data,
        symbolic x coords={
            {24s},{30s},{40s},{50s},{60s},{70s},{80s},{90s}},
            x tick label style={font=\small},
            y tick label style={font=\small},
            width=\linewidth,
            height={0.4\linewidth},
            grid=both,
            grid style={dashed},
            major grid style={dashed},
            tick label style={font=\small},
            label style={font=\small},
            title style={font=\small},
            legend style={at={(1,1.3)}, legend columns=2}
        ]
    \addplot[magenta, mark=*] coordinates {
        ({24s}, 0.652)
        ({30s}, 0.652)
        ({40s}, 0.643)
        ({50s}, 0.639)
        ({60s}, 0.630)
        ({70s}, 0.624)
        ({80s}, 0.621)
        ({90s}, 0.619)
    };
    \addlegendentry{HR.5F};
    \addplot[yellow, mark=*] coordinates {
        ({24s}, 0.753)
        ({30s}, 0.755)
        ({40s}, 0.748)
        ({50s}, 0.739)
        ({60s}, 0.737)
        ({70s}, 0.736)
        ({80s}, 0.732)
        ({90s}, 0.725)
    };
        \addlegendentry{HR3F};
        \end{axis}
        \end{tikzpicture}
        \label{fig:down-sampling_hr_1}
    \end{subfigure}
    
    \caption{Effect of fine-tuning audio window length. Longer windows generally lead to higher ACC, although gains diminish beyond 50s.}
    \label{fig:length_ablation}
\end{figure}
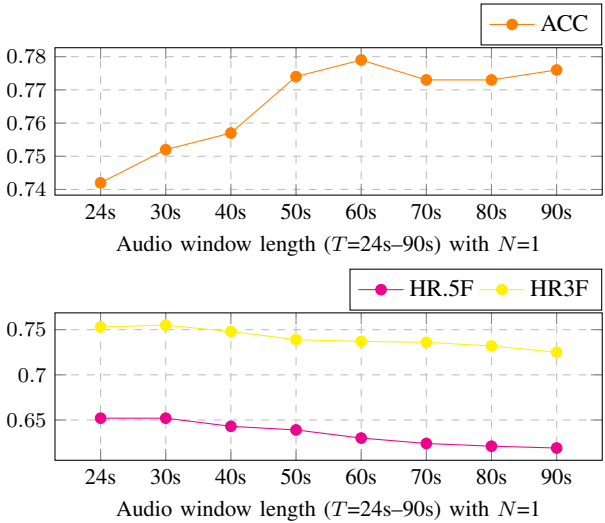

To assess the impact of audio window length, we conduct an ablation study. All ablation studies are conducted on one fixed pre-split Harmonix Set to reduce computation cost. Specifically, we compare the performance when the fine-tuning window length is shorter (24s), equal to (30s), or longer (40--90s) than the pre-training window length (30s). The 90-second limit represents the maximum input length that can be processed on a single A100-80G GPU with a batch size of 8. Figure~\ref{fig:length_ablation} shows the results.
We observe that ACC generally improves as the audio window length increases, although gains become marginal beyond 50 seconds. In contrast, both HR.5F and HR3F decline with longer inputs, in line with prior findings~\cite{tocatch}. These results suggest that while longer audio windows initially help the model capture long-term features and section transitions, the benefits diminish due to a growing mismatch between pre-training and fine-tuning conditions. This transfer learning gap ultimately limits performance as input length increases further.

\subsubsection{Down-sampling Ratio}\label{sec:down-sampling}

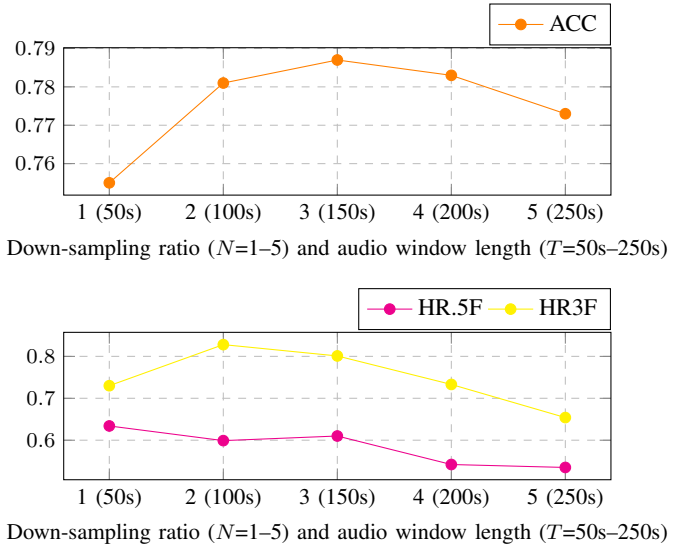
\begin{figure}[tbp]
    \centering
    \begin{subfigure}
        \centering
        \begin{tikzpicture}
        \begin{axis}[
            xlabel={Down-sampling ratio ($N$=1--5) and audio window length ($T$=50s--250s)},
            xtick=data,
            symbolic x coords={{1 (50s)},{2 (100s)},{3 (150s)},{4 (200s)},{5 (250s)}},
            x tick label style={font=\small},
            y tick label style={font=\small},
            width=\linewidth,
            height={0.4\linewidth},
            grid=both,
            grid style={dashed},
            major grid style={dashed},
            tick label style={font=\small},
            label style={font=\small},
            title style={font=\small},
            legend style={at={(1,1.3)}, legend columns=1}
        ]
        \addplot[orange, mark=*] coordinates {
            ({1 (50s)}, 0.755)
            ({2 (100s)}, 0.781)
            ({3 (150s)}, 0.787)
            ({4 (200s)}, 0.783)
            ({5 (250s)}, 0.773)
        };
        \addlegendentry{ACC};
        \end{axis}
        \end{tikzpicture}
        \label{fig:down-sampling_acc_2}
    \end{subfigure}
    
    \vskip\baselineskip
    \vspace{-0.5cm}
    
    \begin{subfigure}
        \centering
        \begin{tikzpicture}
        \begin{axis}[
            xlabel={Down-sampling ratio ($N$=1--5) and audio window length ($T$=50s--250s)},
            xtick=data,
            symbolic x coords={{1 (50s)}, {2 (100s)}, {3 (150s)}, {4 (200s)}, {5 (250s)}},
            x tick label style={font=\small},
            y tick label style={font=\small},
            width=\linewidth,
            height={0.4\linewidth},
            grid=both,
            grid style={dashed},
            major grid style={dashed},
            tick label style={font=\small},
            label style={font=\small},
            title style={font=\small},
            legend style={at={(1,1.3)}, legend columns=2}
        ]
        \addplot[magenta, mark=*] coordinates {
            ({1 (50s)}, 0.634)
            ({2 (100s)}, 0.599)
            ({3 (150s)}, 0.610)
            ({4 (200s)}, 0.542)
            ({5 (250s)}, 0.535)
        };
        \addlegendentry{HR.5F};
        \addplot[yellow, mark=*] coordinates {
            ({1 (50s)}, 0.73)
            ({2 (100s)}, 0.828)
            ({3 (150s)}, 0.801)
            ({4 (200s)}, 0.733)
            ({5 (250s)}, 0.654)
        };
        \addlegendentry{HR3F};
        \end{axis}
        \end{tikzpicture}
        \label{fig:down-sampling_hr_2}
    \end{subfigure}
    
    \caption{Effect of down-sampling ratio. Moderate down-sampling (e.g., 2 or 3) maintains high ACC, but higher ratios lead to HR drops.}
    \label{fig:down-sampling_ablation}
\end{figure}

Figure~\ref{fig:down-sampling_ablation} illustrates the trade-off between function accuracy and boundary hit rate across different down-sampling ratios. When the ratio is below 3, ACC improves steadily, while HR.5F and HR3F remain relatively stable. However, beyond this point, boundary hit rates drop significantly due to the loss of temporal resolution, which is critical for precise boundary detection. For example, at lower ratios, HR.5F allows a tolerance of up to 4 frames, but with higher ratios, even a one-frame shift can result in a miss. Although a larger down-sampling ratio causes sharper boundary peaks, the peak picking performance remains similar under different ratios. Overall, down-sampling ratios of 2 or 3 achieve the best balance between function accuracy and boundary precision. 


\subsubsection{Contrastive Learning}
\begin{figure}[htbp]
        \centering
        \begin{tikzpicture}
        \begin{axis}[
            ybar,
            xlabel={Audio window length ($T$=60s--180s) with $N$=2},
            ylabel={Accuracy (ACC)},
            xtick=data,
            symbolic x coords={60s ($N$=2), 100s ($N$=2), 180s ($N$=2)},
            x tick label style={font=\small},
            y tick label style={font=\small},
            width=0.9\linewidth,
            height={0.35\linewidth},
            grid=both,
            grid style={dashed},
            bar width=0.25cm,
            legend style={at={(0.5,1.1)}, anchor=south, legend columns=2},
        ]

        \addplot[fill=magenta!50, xshift=0cm] coordinates {
            (60s ($N$=2), 0.784)
            (100s ($N$=2), 0.787)
            (180s ($N$=2), 0.783)
        };
        \addlegendentry{MLP w. CL};

        \addplot[fill=blue!50, xshift=0cm] coordinates {
            (60s ($N$=2), 0.751)
            (100s ($N$=2), 0.802)
            (180s ($N$=2), 0.801)
        };
        \addlegendentry{Transformer w/o. CL};

        \addplot[fill=red!50, xshift=0cm] coordinates {
            (60s ($N$=2), 0.763)
            (100s ($N$=2), 0.814)
            (180s ($N$=2), 0.805)
        };
        \addlegendentry{Transformer w. CL};
        \end{axis}
        \end{tikzpicture}

    \caption{Ablation study for contrastive learning and probing heads.}
    \label{fig:cl}
\end{figure}
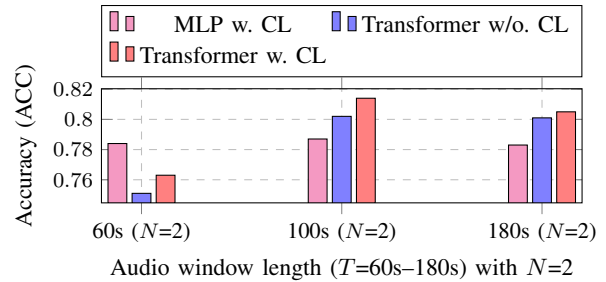

Figure~\ref{fig:cl} compares the ACC scores of models trained with and without contrastive learning. The results show a consistent, marginal improvement in ACC across all tested models when contrastive learning is involved. This enhancement contributes to the overall performance gains observed in our main results, helping the model achieve new state-of-the-art results.


\begin{table}[tbp]
\footnotesize
    \centering
    \begin{tabular}{c|c|cc}
    \toprule
    Length ($T$) & Ratio ($N$)    & Time/Batch & GPU Memory \\
         \midrule
     50s & 1& 0.65s &  48.36\% \\
     100s &2  & 0.65s & 49.06\%  \\
     \midrule
     90s & 1 & 1.25s & 98.69\%  \\
     180s & 2  & 1.25s &  98.76\% \\
         \bottomrule
    \end{tabular}
    \caption{Comparison of computational cost.}
    \label{tab:my_label}
\end{table}

\subsubsection{Computing Cost}
Table~\ref{tab:my_label} compares the computational cost associated with different audio window lengths and down-sampling ratios. Our result shows that extending the audio window length ($T$) results in a roughly linear increase in GPU memory usage. In contrast, increasing the down-sampling ratio ($N$) has minimal impact on GPU memory and computation time, while still has performance gains. 

\subsubsection{Probing Head}
Table~\ref{tab:my_label} shows that both MLP and Transformer Heads achieve SOTA performance. The MLP head performs better at shorter window lengths but does not improve with increased length, suggesting the Transformer head is superior for mitigating the adaptation gap with longer fine-tuning lengths relative to pretraining.




\section{Conclusion}
This work has demonstrated that temporal adaptation for pre-trained music foundation models offers an effective and computationally efficient approach to MSA. By fine-tuning with extended audio windows and reduced temporal resolution, we significantly expand the models’ receptive fields, enabling them to capture the full context of a song while maintaining consistent structural predictions.

Extensive experiments show that proper down-sampling combined with longer input durations enhances both structural function prediction and boundary detection, achieving state-of-the-art performance on benchmark datasets. These findings suggest that relaxing the temporal constraints imposed during pre-training can improve generalization and scalability, opening new directions for research in MSA and other MIR tasks.




\clearpage
\bibliographystyle{IEEEtran}
\bibliography{refs25}







\end{document}